\documentclass[
    aps,
    amsmath,amssymb,
    superscriptaddress
    ]{revtex4-2}
\usepackage{siunitx}
\usepackage{graphicx}
\newcommand{\Sun}[0]{\odot}
\usepackage{newtxtext}
\usepackage{newtxmath}
\usepackage{xcolor}
\usepackage{hyperref}

\begin{document}

\title{TOrsion-Bar Antenna: A Ground-Based Detector
for Low-Frequency Gravity Gradient Measurement}

\author{Satoru Takano}
\email[Correspondence: ]{takano@gw.phys.titech.ac.jp}
\affiliation{Department of Physics, The University of Tokyo, Bunkyo-ku, Tokyo 133-0033, Japan}
\affiliation{Department of Physics, Tokyo Institute of Technology, Meguro-Ku, Tokyo 152-8551, Japan}

\author{Tomofumi Shimoda}
\altaffiliation[Current address: ]{National Metrology Institute of Japan (NMIJ), National Institute of Advanced Industrial
Science and Technology (AIST), Tsukuba, Ibaraki 305-8563, Japan}
\affiliation{Department of Physics, The University of Tokyo, Bunkyo-ku, Tokyo 133-0033, Japan}

\author{Yuka Oshima}
\affiliation{Department of Physics, The University of Tokyo, Bunkyo-ku, Tokyo 133-0033, Japan}

\author{Ching Pin Ooi}
\affiliation{Department of Physics, The University of Tokyo, Bunkyo-ku, Tokyo 133-0033, Japan}

\author{Perry William Fox Forsyth}
\affiliation{Department of Physics, The University of Tokyo, Bunkyo-ku, Tokyo 133-0033, Japan}

\author{Mengdi Cao}
\affiliation{Department of Astronomy, Beijing Normal University, Beijing 100875, China}

\author{Kentaro Komori}
\affiliation{Department of Physics, The University of Tokyo, Bunkyo-ku, Tokyo 133-0033, Japan}
\affiliation{Research Center for the Early Universe (RESCEU), Graduate School of Science, The University of Tokyo, Bunkyo-ku, Tokyo 113-0033, Japan}

\author{Yuta Michimura}
\affiliation{Research Center for the Early Universe (RESCEU), Graduate School of Science, The University of Tokyo, Bunkyo-ku, Tokyo 113-0033, Japan}

\author{Ryosuke Sugimoto}
\affiliation{Department of Physics, The University of Tokyo, Bunkyo-ku, Tokyo 133-0033, Japan}

\author{Nobuki Kame}
\affiliation{Earthquake Research Institute, The University of Tokyo, Bunkyo-ku, Tokyo 113-0032, Japan}

\author{Shingo Watada}
\affiliation{Earthquake Research Institute, The University of Tokyo, Bunkyo-ku, Tokyo 113-0032, Japan}

\author{Takaaki Yokozawa}
\affiliation{Institute for Cosmic Ray Research (ICRR), KAGRA Observatory, The University of Tokyo, Kashiwa 277-8582, Japan}

\author{Shinji Miyoki}
\affiliation{Institute for Cosmic Ray Research (ICRR), KAGRA Observatory, The University of Tokyo, Kashiwa 277-8582, Japan}

\author{Tatsuki Washimi}
\affiliation{Gravitational Wave Science Project (GWSP), Kamioka Branch, National Astronomical Observatory of Japan
(NAOJ), Kamioka-cho, Hida City, Gifu 506-1205, Japan}

\author{Masaki Ando}
\affiliation{Department of Physics, The University of Tokyo, Bunkyo-ku, Tokyo 133-0033, Japan}
\affiliation{Research Center for the Early Universe (RESCEU), Graduate School of Science, The University of Tokyo, Bunkyo-ku, Tokyo 113-0033, Japan}

\date{\today}

\begin{abstract}
The Torsion-Bar Antenna (TOBA) is
a torsion pendulum-based gravitational detector
developed to observe gravitational waves
in frequencies between
$\SI{1}{mHz}$ and $\SI{10}{Hz}$.
The low resonant frequency of the torsion pendulum
enables observation in this frequency band on the ground.
The final target of TOBA is
to observe gravitational waves with a 10 m detector
and expand the observation band of gravitational waves.
In this paper, an overview of TOBA, including the previous prototype experiments and the current ongoing development, is presented.
\end{abstract}

\maketitle

\section{\label{sec:1}Introduction}

Since the first detection of gravitational waves (GWs)
by Advanced LIGO \cite{150914},
gravitational wave astronomy has continued to develop
and multi-messenger astronomy has gained attention.
The sources of the detected GWs
by the ground-based detectors
are the coalescence of binary systems
with source masses in the range of
1 $M_\Sun$--100 $M_\Sun$,
which correspond to the observed frequency band of
$\SI{10}{Hz}$--$\SI{1}{kHz}$ \cite{GWTC3}.
Another observation band of interest is
in the range of $\SI{1}{mHz}$--$\SI{10}{Hz}$.
In this frequency band,
some target sources are expected,
such as the merger of
Intermediate Mass Black Holes (IMBHs)
with masses in the order of
10$^{2}$ $M_\Sun$--10$^{5}$ $M_\Sun$,
or the Stochastic Gravitational Wave Background
(SGWB) from the early universe.
To observe GWs in this frequency band,
space interferometric detectors such as LISA \cite{LISA} and DECIGO \cite{DECIGO}
have been proposed.

\begin{sloppypar}
    Several ground-based detectors
    for observing GWs in frequencies between $\SI{1}{mHz}$ and $\SI{10}{Hz}$
    have been proposed,
    such as atomic interferometers \cite{ZAIGA, MIGA}
    and super-conductive gravity gradiometers
    \cite{SGG3D, SOGRO}.
    One such detector is the Torsion-Bar Antenna
    (TOBA),
    which uses torsion pendulums to observe GWs
    \cite{TOBA}.
    Thanks to the low resonant frequency of the torsion pendulums,
    TOBA is sensitive to the gravitational waves in
    the frequency band of $\SI{1}{mHz}$--$\SI{10}{Hz}$.
    The main goal of TOBA is to achieve the sensitivity
    of $10^{-19\,}/\si{\sqrt{Hz}}$ at \mbox{0.1 Hz}.
    This sensitivity allows us to detect IMBH binary mergers
    at cosmological distances.
    Thanks to the ground-based configuration,
    TOBA has the advantage of better accessibility
    and lower cost than the space detectors
    in the similar frequency band.
\end{sloppypar}

In this paper, we provide an overview of TOBA.
The principle of operation and scientific targets of the TOBA
are in Section \ref{sec:2}.
Next, the results of the previous prototypes
and the related developments are explained in Section \ref{sec:3}.
In Section \ref{sec:4}, we introduce the upcoming prototype,
Phase-III TOBA, and present recent developments.

\section{\label{sec:2}Overview of TOBA}

\subsection{\label{sec:2.1}Principle of TOBA}

The theory guiding the usage of torsion pendulums
to detect GWs predates the TOBA detector
by several decades \cite{Pre-TOBA1,Pre-TOBA2}.
A detailed derivation of the principal of detection
is provided in Appendix \ref{sec:A}
but summarized here.

Figure \ref{TOBA} shows a basic configuration of TOBA,
with bars suspended in the horizontal $xy$-plane.
Here, we focus on the rotational motion of a bar
around the $z$-axis, $\theta$.

\begin{figure}[btp]
    \centering
	\includegraphics[width=1\linewidth]
    {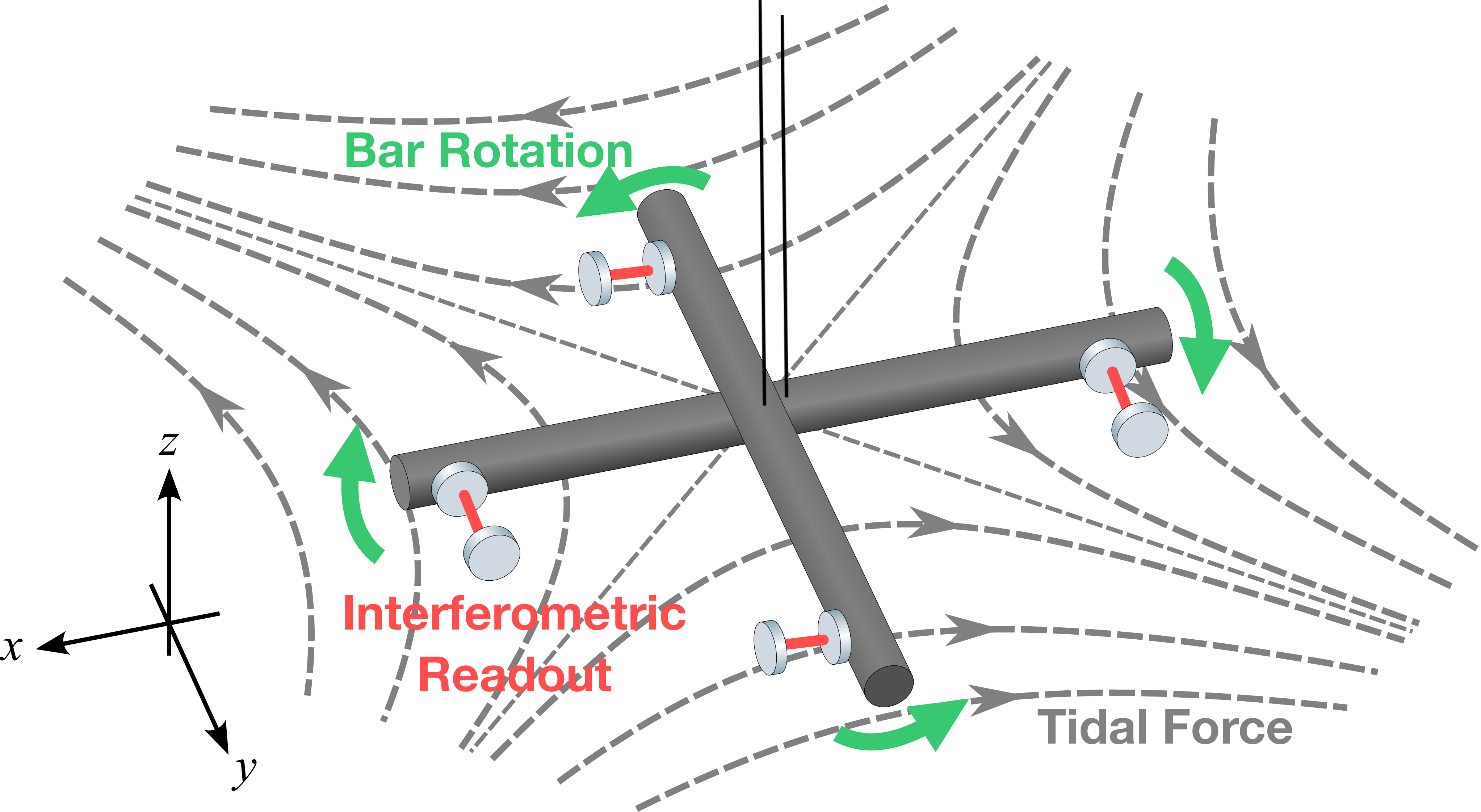}
	\caption{\label{TOBA}
    Principle of TOBA.
    Two test mass bars are suspended
    in the horizontal plane
    independently and orthogonally.
    Dashed lines indicate the gravitational tidal forces
    induced by a GW coming along the $z$-axis,
    and green arrows show the rotation directions
    of the bars.
    The rotation angles of the bars are measured
    by interferometers.}
\end{figure}

Suppose that a GW comes along the $z$-axis,
\begin{equation}
	h_{ij}^{A} = h_Ae_{ij}^{A}\cos 2\pi ft,
    \label{GW_z}
\end{equation}
where
\begin{equation}
	e_{ij}^+ = \begin{pmatrix}
		1 & 0 & 0 \\
		0 & -1 & 0 \\
		0 & 0 & 0
	\end{pmatrix},\,
	e_{ij}^\times = \begin{pmatrix}
		0 & 1 & 0 \\
		1 & 0 & 0 \\
		0 & 0 & 0
	\end{pmatrix}
    \label{GW_pol}
\end{equation}
are the polarization tensors of GW.
Considering the forces acting on each discrete element of the bar,
the frequency response function is given by
\begin{equation}
	H_+(f) := \frac{\theta}{h_+} = 0,\,\,
    H_\times(f) := \frac{\theta}{h_\times}
	\simeq \frac{1}{2}
    \frac{f^2}{f_0^2(1+i\phi_\mathrm{rot})-f^2},
\end{equation}
where $f_0$ is the resonant frequency of the torsion pendulum
and $\phi_\mathrm{rot}$ is the loss angle of the suspension wire.

The differential response of each bar
is generated by the perpendicular orientation
of each bar.
By matching mechanical parameters
between the two pendulums,
such as overlapping the center of mass and
maintaining a similar mechanical response,
noises can be rejected from the GW measurement
provided by the differential motion of the bars.
This rejection of common noises
between the two pendulums is referred to
as common mode rejection of the detector,
and stands as one of the significant advantages
of using a twin torsional pendulum detector setup
for GW detection, such as the TOBA.

\subsection{\label{sec:2.2}Target Sensitivity of TOBA}

\begin{figure}[btp]
    \centering
	\includegraphics[width=1\linewidth]{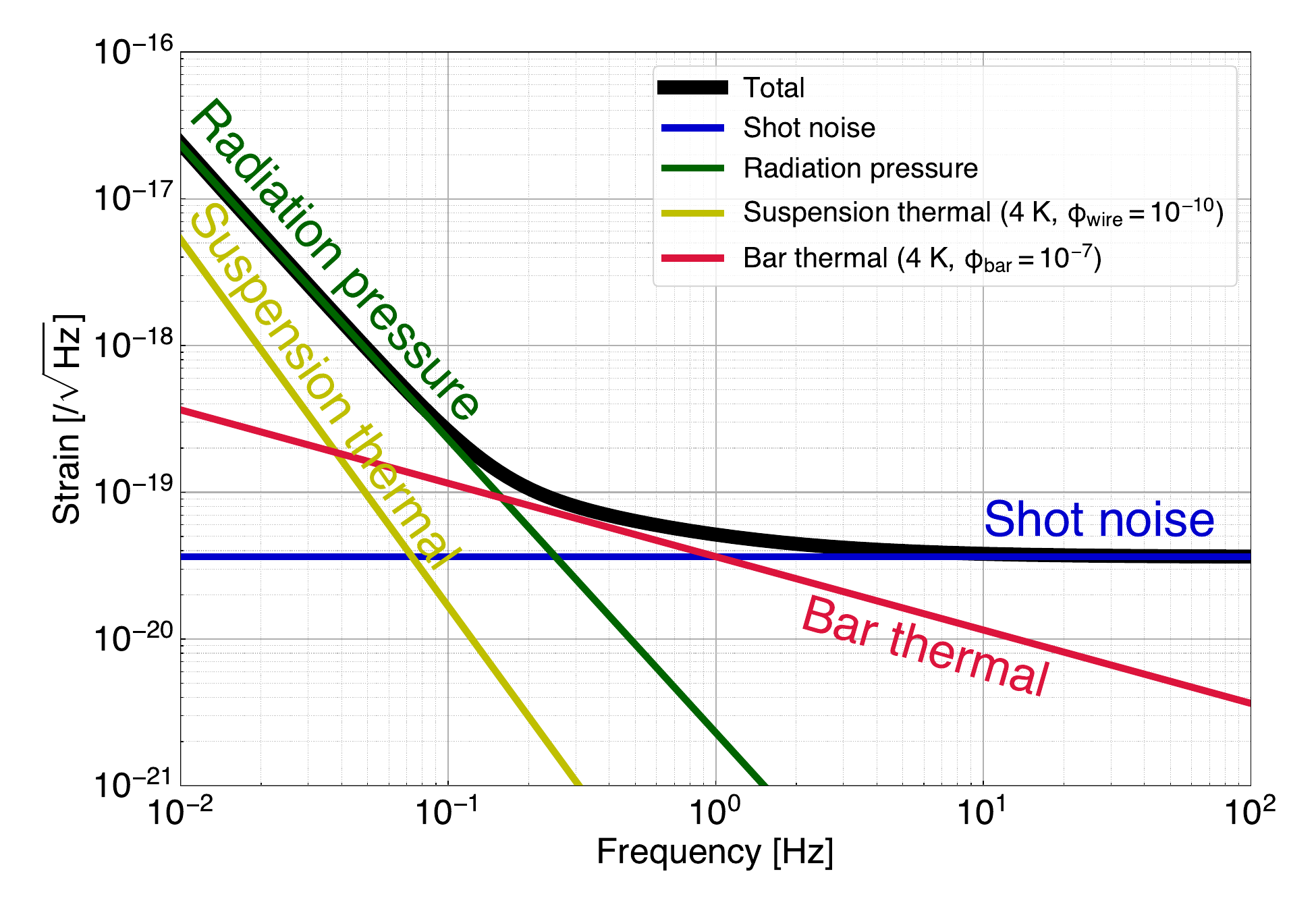}
	\caption{\label{FinalTOBA}
    The design sensitivity of the Final TOBA \cite{TOBA}.
    The blue and green lines show the photon shot noise
    and radiation pressure noise, respectively.
    The yellow and red lines indicate the thermal noise
    of the suspension wire and the test mass bars.
    The black line is the sum of all noises.}
\end{figure}

The final target sensitivity of TOBA is
$10^{-19}\,/\si{\sqrt{Hz}}$ at $\SI{0.1}{Hz}$
in strain,
which is determined
for observing the targets mentioned in the following sections.
To achieve this sensitivity,
a $\SI{10}{m}$ torsion-bar detector,
The Final TOBA, is proposed.
The designed noise budget is shown in Figure \ref{FinalTOBA}.

The Final TOBA consists of two large torsion pendulums, as shown in Figure \ref{TOBA}.
Each bar is made of aluminum
with a length of $\SI{10}{m}$ and
a diameter of $\SI{0.6}{m}$,
which correspond to
a mass of $\SI{7.6e3}{kg}$ and
moment of inertia of $\SI{6.4e4}{kg.m^2}$.
The resonant frequency of the rotational mode
is designed to be $\SI{1}{mHz}$
and the loss of the suspension wire
and the loss of the internal modes of the bar
are required to be
$10^{-10}$ and $10^{-7}$, respectively.
Each bar and its suspension wire are to be kept
at a cryogenic temperature of $\SI{4}{K}$
to reduce thermal noise.
The rotation of the bars is to be measured by
a pair of Fabry--Pérot cavities
at both ends of the bars.
Each cavity has a finesse of 100 and
the input laser power to the cavities
is to be set as $\SI{10}{W}$.
Two pendulums are to have as a close mechanical response
as possible, which can eliminate or reduce noises
familiar to both pendulums.

Considering all these noises,
the sensitivity of the Final TOBA is predicted to be
limited by the quantum noise
(shot noise and radiation pressure noise)
and thermal noise of the bar \cite{TOBA}.

\subsection{\label{sec:2.3}
Target Sources of Gravitational Waves for TOBA}

One of the main target sources of GWs
in the frequency band of $\SI{1}{mHz}$--$\SI{10}{Hz}$
is the coalescence of IMBH binary systems.
Their existence has only been reported through indirect mass estimation, and direct observation of GWs from these systems is of great importance
for the study of IMBHs.
For example, observation of GWs from IMBHs
would provide us with some hints
about the origin and formation
of supermassive black holes
at the center of galaxies \cite{IMBH1,IMBH2}.
With the Final TOBA's target sensitivity,
the observation range of IMBH binary mergers
with their source mass $10^5$ $M\Sun$ reaches $\SI{10}{Gpc}$,
which corresponds to
the redshift of $z$ $\sim$ $2.4$ \cite{TOBA}.

Another important target of GWs
in this frequency band is
the SGWB from the early universe.
Unlike electromagnetic waves,
GWs can pass through the radiation field before recombination
and see the universe before 0.38 Myr from the Big Bang.
In terms of the dimensionless energy density of GWs,
$\Omega_\mathrm{GW}$,
the sensitivity of $10^{-19}\,/\si{\sqrt{Hz}}$ corresponds to $\Omega_\mathrm{GW} \simeq 10^{-7}$
with a one-year observation period \cite{TOBA},
which is a better constraint than
the upper bound set by
Big Bang Nucleosynthesis (BBN) \cite{BBN}.

\subsection{\label{sec:2.4}Geophysical Target of TOBA}

Since GW detectors respond to the gravity gradient,
the measured signal can also originate from
terrestrial sources such as the ground and the atmosphere.
In terms of GW detection, such a signal is called Newtonian Noise (NN)
because the gravity gradient induced by the terrestrial sources
cannot be distinguished from the effect of GWs \cite{NN}.
It has also been proposed that
with a moderate sensitivity of $10^{-15}\,/\si{\sqrt{Hz}}$,
the measured terrestrial gravity gradient has
great geophysical importance \cite{shimoda_warning}.
Here, we introduce two related cases.

\subsubsection{\label{sec:2.4.1}
Newtonian Noise Measurement}

NN is a fundamental noise source
for GW detectors.
For future ground-based GW detectors,
NN is assumed to be the dominant noise source
in frequency bands below 10 Hz \cite{ET,CE},
and its mitigation method is discussed in
\cite{NN_sub1,NN_sub2}.
Several models describe the mechanism
of NN generation \cite{NN_seis_LIGO,NN_KAGRA,NN_air_LIGO, NN_cooler}.
However, so far, NN has not been measured directly.
Therefore, direct measurement of NN
and test of the models
are essential for GW detectors on the ground.

Some models predict that around 0.1 Hz,
the amplitude level of NN grows up to
the order of $10^{-15}\,/\si{\sqrt{Hz}}$
\cite{NN_KAGRA,TorPeDO}.
Therefore, TOBA is expected to detect NN in this frequency band
and partially justify the model and cancellation scheme.

\subsubsection{\label{sec:2.4.2}
Early Earthquake Detection}

It is proposed that
transient changes in gravitational fields
introduced by large earthquakes
can be detected by gravimeters
and gravity gradiometers \cite{earthquake_warning}.
This scheme pays attention to the change
that occurs on the timescale of
$\sim$$\SI{10}{s}$--$\SI{100}{s}$,
which corresponds to
the frequency band of TOBA.
Because the change in the gravitational field
propagates with the speed of light, the
detection of earthquakes with TOBA is expected to be
faster than the current warning scheme,
which uses seismic P-waves with a velocity of
$\sim$$\SI{1}{km/s}$--$\SI{10}{km/s}$.

It has also been pointed out that
TOBA has an advantage in detecting earthquakes.
Gravimetric measurements,
such as using gravimeters, seismometers, and tiltmeters,
are considered to be ineffective for early earthquake detection
due to decreased sensitivity caused by instrument acceleration
\cite{earthquake_seis1, earthquake_seis2}.
On the other hand, TOBA is expected to avoid this cancellation
because instrument acceleration does not affect
gravity gradient measurements.

In previous research,
the possibility of detecting earthquakes
with TOBA was \mbox{studied \cite{shimoda_warning,Shimoda_PhD}}.
It is expected that
with a moderate sensitivity of $\sim$$10^{-15}\,/\si{\sqrt{Hz}}$
at $\SI{0.1}{Hz}$,
the detectability is almost the same as the current system.

\section{\label{sec:3}Prototype Developments}

\begin{figure}[btp]
    \centering
	\includegraphics[width=0.90\linewidth]{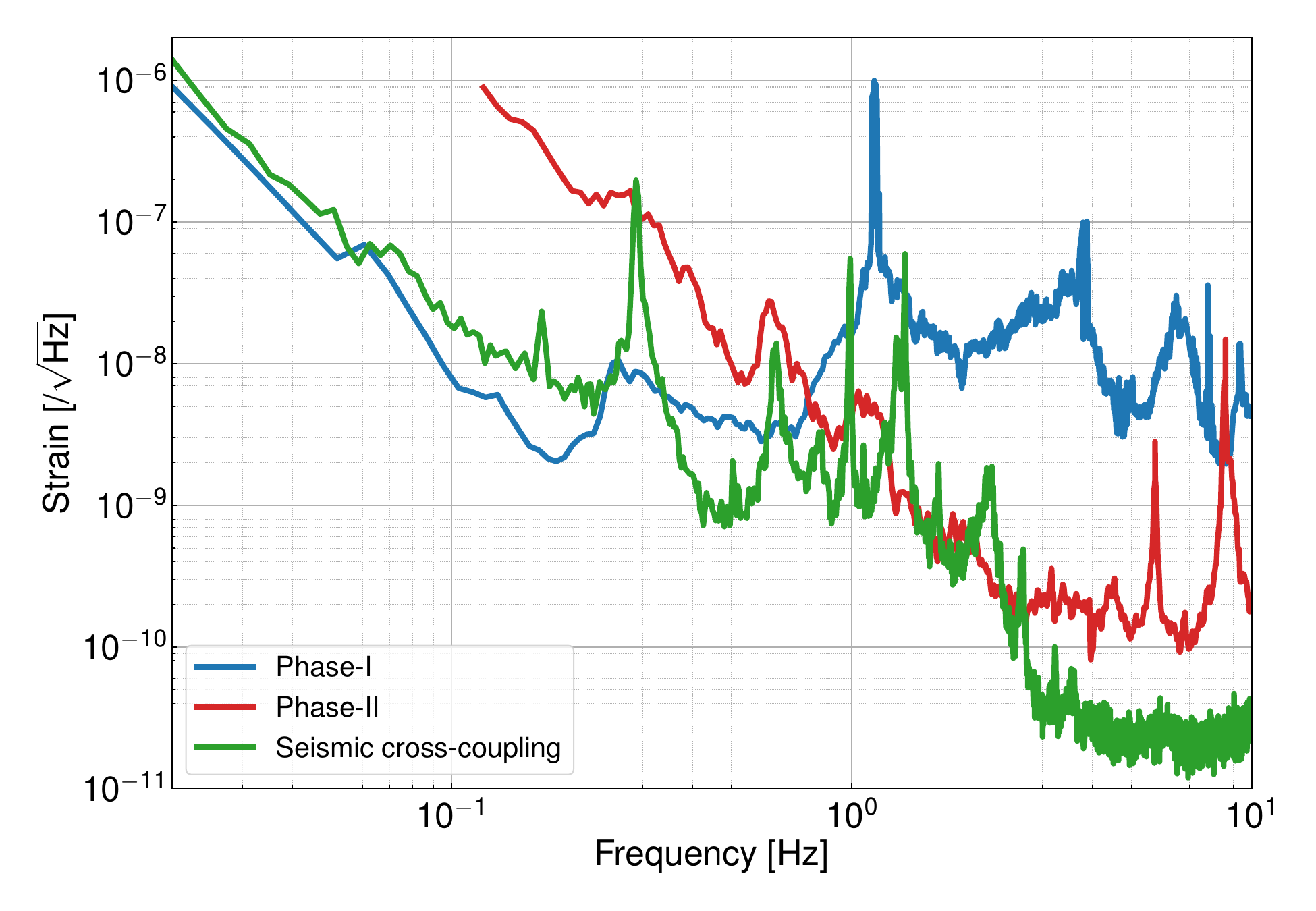}
	\caption{\label{Sens_proto}
    The sensitivity of prototype experiments.
    The blue and red curves are the achieved sensitivities
    of Phase-I \cite{GWB_phase1} and Phase-II \cite{shoda}.
    The green curve shows the sensitivity obtained
    in the experiment of seismic cross-coupling
    reduction \cite{phase2-5}.}
\end{figure}

To reach the target sensitivity of the Final TOBA,
studies of each component
and demonstrations of noise reduction have been performed
with small prototypes.
The achieved sensitivity of these prototypes are
summarized in Figure \ref{Sens_proto}.

In this section, we briefly introduce these
prototype developments and their achievements.

\subsection{\label{sec:3.1}Phase-I TOBA}

\begin{figure}[btp]
    \centering
    \begin{tabular}{cc}
        \begin{minipage}{.35\textwidth}
            \centering
            \includegraphics[width=0.7\linewidth]{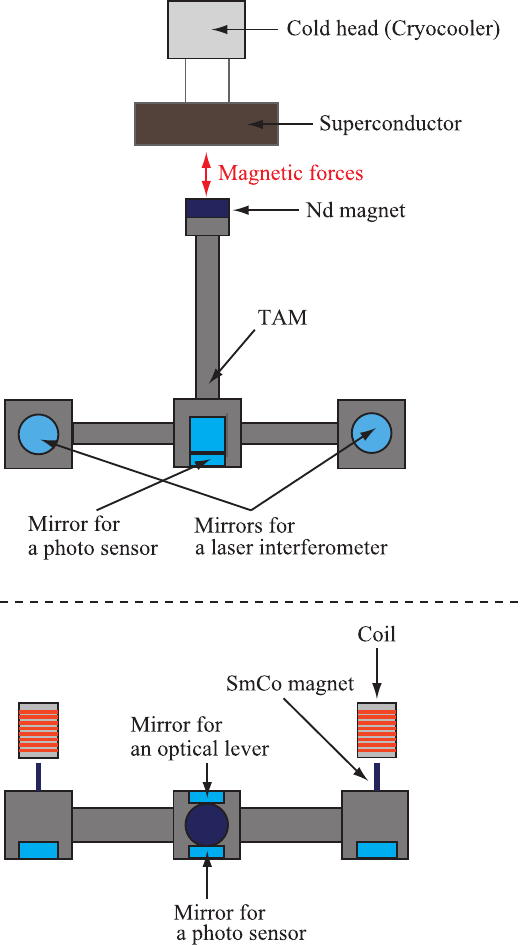}
        \end{minipage}
        \begin{minipage}{.45\textwidth}
            \centering
            \includegraphics[width=0.9\linewidth]{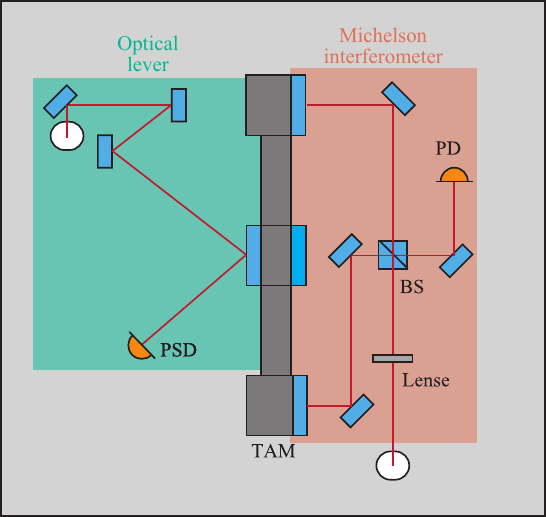}
        \end{minipage}
    \end{tabular}
    \caption{\label{phase1_fig}
    The configuration of Phase-I TOBA.
    (\textbf{Left}) Test mass bar of Phase-I TOBA.
    A single test mass bar was suspended by
    superconducting magnet levitation.
    The test mass was $\SI{22.5}{cm}$ bar
    made of aluminum. Many mirrors were attached
    to monitor the motion.
    (\textbf{Right}) The optical configuration of Phase-I TOBA.
    The best sensitivity was obtained with a Michelson
    interferometer consisting of the end mirrors
    attached to both ends of the bar
    and the other optics on an optical bench
    fixed on the vacuum chamber.
    These figures were adapted from \cite{ishidoshiro}
    and modified for this paper.
    }
\end{figure}

The first prototype, named Phase-I TOBA,
was developed in 2010
\cite{ishidoshiro}
as a proof of concept.
The test mass bar, with a length of $\SI{22.5}{cm}$,
was suspended by superconducting
magnet levitation.
Though there is no restoring force
in the rotational degree of freedom in principle,
the actual resonant frequency was $\sim$$\SI{5}{mHz}$
due to the gradient of the magnetic field
induced by the levitating magnet.
The rotational motion was measured by
a Michelson interferometer on the ground.
The configuration of the bar and the interferometer
is shown in Figure \ref{phase1_fig}.

The sensitivity of the Phase-I prototype was limited by
ambient magnetic field fluctuations and seismic cross-coupling.
Below $\SI{0.1}{Hz}$, the magnet for levitating the test mass bar
coupled with the ambient magnetic field
and induced a torque noise to the test mass bar.
Above $\SI{0.1}{Hz}$,
the translational seismic motion
was transferred to the rotational signal
of the test mass bar
due to the relative tilt of the mirrors attached
on the test mass
for interferometric readout.
The achieved sensitivity was
$\sim$$10^{-8}\,/\si{\sqrt{Hz}}$ at $\SI{0.1}{Hz}$.

With this sensitivity,
a search for SGWB was performed, and the upper limit was set to
$\Omega_\mathrm{GW}h_0^2<\num{1.9e17}$
at $\SI{0.035}{Hz}$--$\SI{0.830}{Hz}$
\cite{ishidoshiro, GWB_phase1}.

\subsection{\label{sec:3.2}Phase-II TOBA}

The second prototype, Phase-II TOBA, was developed
in 2015 \cite{shoda,GWB_phase2}.
In this prototype, two test mass bars
with a length of $\SI{24}{cm}$ were used,
and they were suspended by metal wires
in order to reduce magnetic noise coupling,
which was one of the dominant noise sources
in Phase-I TOBA.
Two test mass bars were suspended orthogonally
with double wires from an intermediate mass
and consisted of double-stage pendulums.
At the suspension point, an active feedback system
was introduced for reducing seismic vibration.
It consisted of six seismometers
and hexapod actuators.

\begin{figure}[btp]
    \centering
    \begin{tabular}{cc}
        \begin{minipage}{.52\textwidth}
            \centering
            \includegraphics[width=0.9\linewidth]{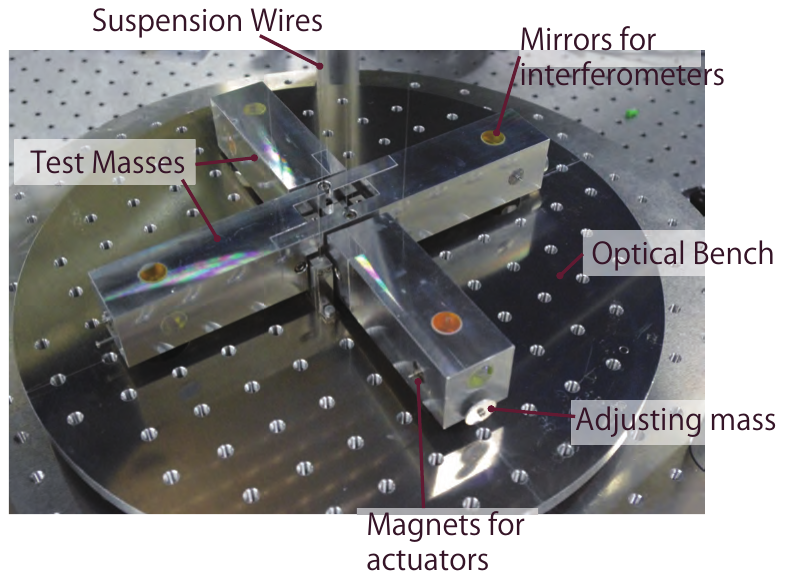}
        \end{minipage}
        \begin{minipage}{.42\textwidth}
            \centering
            \includegraphics[width=0.9\linewidth]{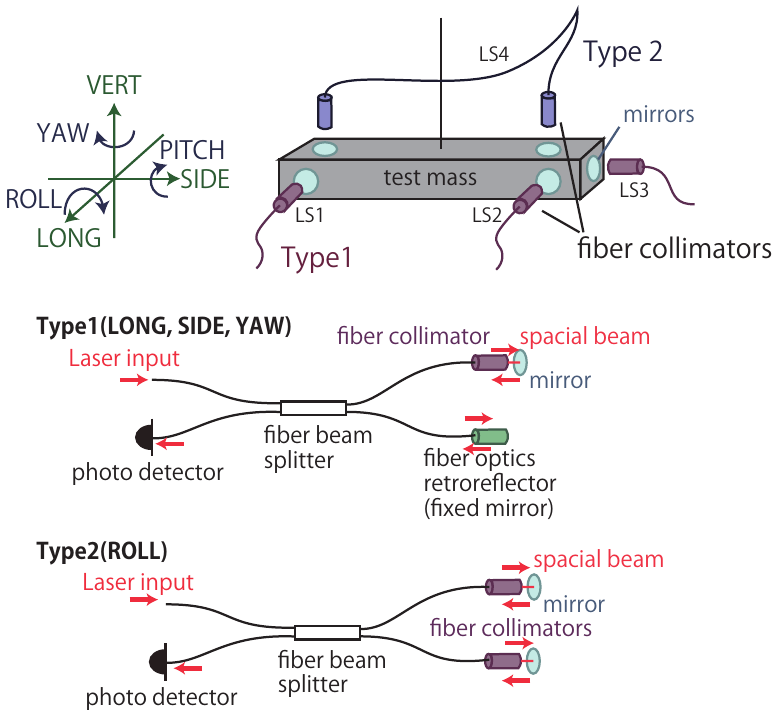}
        \end{minipage}
    \end{tabular}
    \caption{\label{phase2_fig}
    The configuration of Phase-II TOBA.
    (\textbf{Left}) Test mass bars of Phase-II TOBA.
    Two test mass bars were suspended by
    two suspension wires.
    Each test mass was $\SI{24}{cm}$ long and
    made of aluminum. Many mirrors were attached
    for fiber interferometer.
    (\textbf{Right}) The optical configuration of Phase-II TOBA.
    Fiber Michelson interferometers were utilized
    to measure multiple rotation degrees of freedom
    of the test mass bars.
    These figures were adapted from \cite{shoda}.
    }
\end{figure}

The rotational motion was measured by
fiber Michelson interferometers
constructed on an optical bench,
which was also suspended
from the intermediate mass.
With these fiber interferometers,
not only the horizontal rotation
but also the vertical rotational motions
were measured.
Thanks to this configuration called a multi-output configuration,
the detection volume and angular accuracy
can be improved
because the signal from each degree of freedom
compensates for the blind spots
of the other degrees of freedom.
The configuration of the test masses
and interferometric sensors are shown
in Figure \ref{phase2_fig}.

The sensitivity of Phase-II TOBA was limited
by seismic coupling noise and the phase noise
induced by the vibration of the optical fiber
within the readout interferometer.
The achieved sensitivity was
$\sim$$10^{-10}\,/\si{\sqrt{Hz}}$
at $\SI{3}{Hz}$--$\SI{8}{Hz}$.

\begin{sloppypar}
With this sensitivity,
the upper limit on SGWB was set to
$\Omega_\mathrm{GW}h_0^2<\num{1.2e20}$
at \mbox{$\SI{2.6}{Hz}$
\cite{GWB_phase2}}.
Also, the existence of $200$ $M_\Sun$ IMBH binaries
was excluded within \mbox{$r<\SI{2.1e-4}{pc}$
\cite{shoda}.}
\end{sloppypar}

\subsection{\label{sec:3.3}
Research on Seismic Cross-Coupling Reduction}

\begin{figure}[btp]
    \centering
    \begin{tabular}{cc}
        \begin{minipage}{.49\textwidth}
            \centering
            \includegraphics[width=0.95\linewidth]{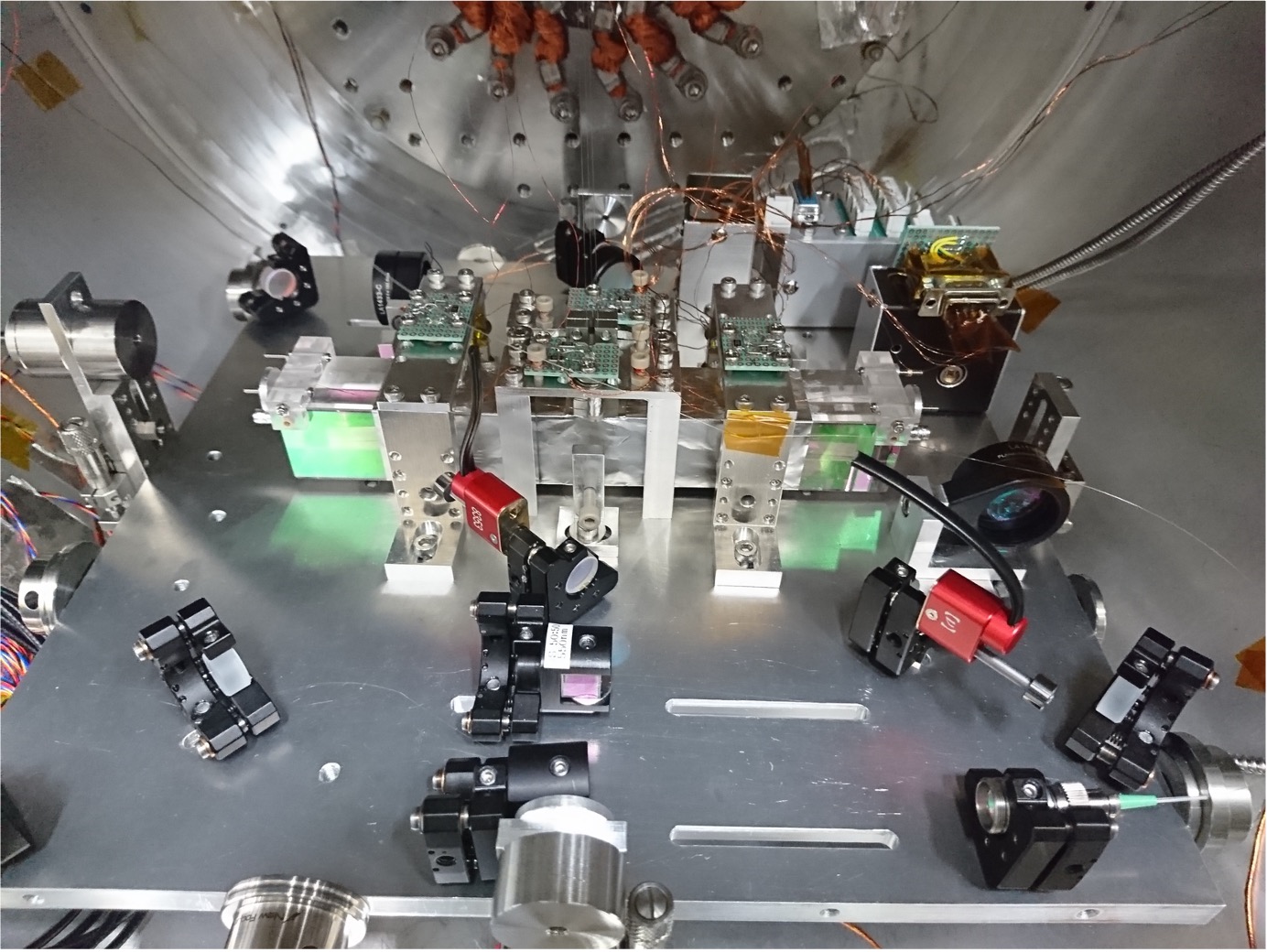}
        \end{minipage}
        \begin{minipage}{.45\textwidth}
            \centering
            \includegraphics[width=0.95\linewidth]{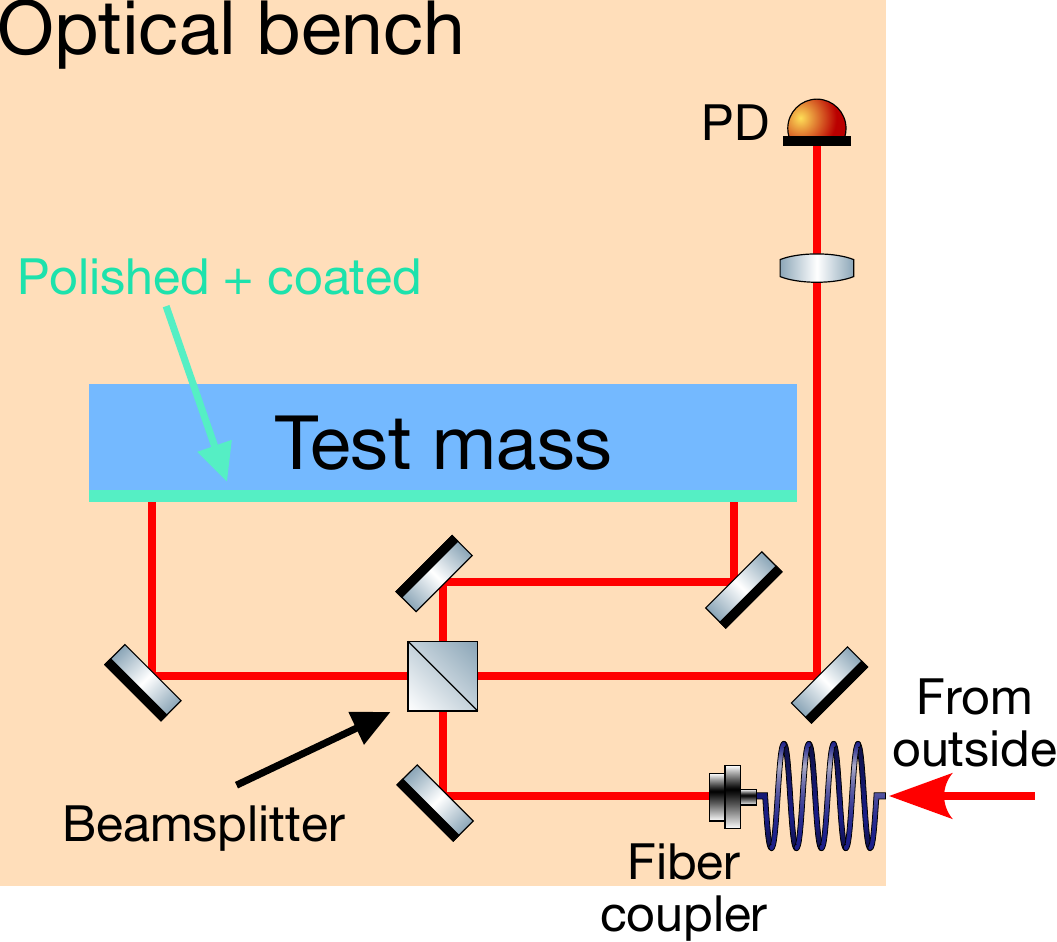}
        \end{minipage}
    \end{tabular}
    \caption{\label{phase2.5_fig}
    Thes etup for research
    on seismic cross-coupling reduction.
    (\textbf{Left}) A picture of
    the test mass bar and the optics.
    The test mass bar was made of fused silica,
    and its front surface was polished and HR-coated.
    The length of the test mass was $\SI{20}{cm}$,
    and it was suspended by a single wire.
    \mbox{(\textbf{Right}) The} optical configuration of this setup.
    A Michelson interferometer consisted of
    the optics on the suspended optical bench
    and the surface of the test mass.
    }
\end{figure}

For understanding the translational
seismic noise coupling
and reducing it,
an experiment with a double-stage pendulum was
performed in 2017 \cite{phase2-5}.
This experiment consisted of a test mass bar
and free-space Michelson interferometer on an optical bench
to measure the horizontal rotation of the bar.
The bar was a $\SI{20}{cm}$ long fused silica block
with an optical coating on a well-polished surface.
By making the interferometer with this flat surface,
the relative tilt between the mirror surface
at both ends of the bar was reduced,
which was the main source of the cross-coupling
in Phase-I TOBA.
The configuration of the test mass
and the interferometer is shown
in Figure \ref{phase2.5_fig}.

This experiment examined the model
of the cross-coupling of the seismic noise,
and demonstrated the reduction scheme of the coupling
by tuning the tilt angle of the test mass bar
and the optical bench.
The achieved cross-coupling function was
$\SI{5e-6}{rad/m}$ at $\SI{0.1}{Hz}$.
Even though these values are not sufficient
for the Final TOBA,
this work established the basic strategy
for reducing the coupling transfer function.

\section{\label{sec:4}Next Prototype: Phase-III TOBA}

Following these prototype experiments,
an upgraded prototype, Phase-III TOBA,
is currently under development.
We consider Phase-III TOBA to be
just one step before the Final TOBA,
and the configuration is closer to it
than the previous experiments.
The target of Phase-III TOBA is
to reach a sensitivity of about $10^{-15}\,/\si{\sqrt{Hz}}$ at $\SI{0.1}{Hz}$
with small-scale ($\sim$$\SI{30}{cm}$) test mass bars.
In addition to the noise reduction mentioned above,
the reduction in the thermal noise
by introducing a cryogenic system is
one of the important issues
to be solved here.
Furthermore, with the target sensitivity,
we expect to achieve scientific results,
such as the detection of earthquakes and
direct measurement of NN.

\subsection{\label{sec:4.1}Design Overview}

\begin{figure}[btp]
    \centering
    \includegraphics[width=1\linewidth]{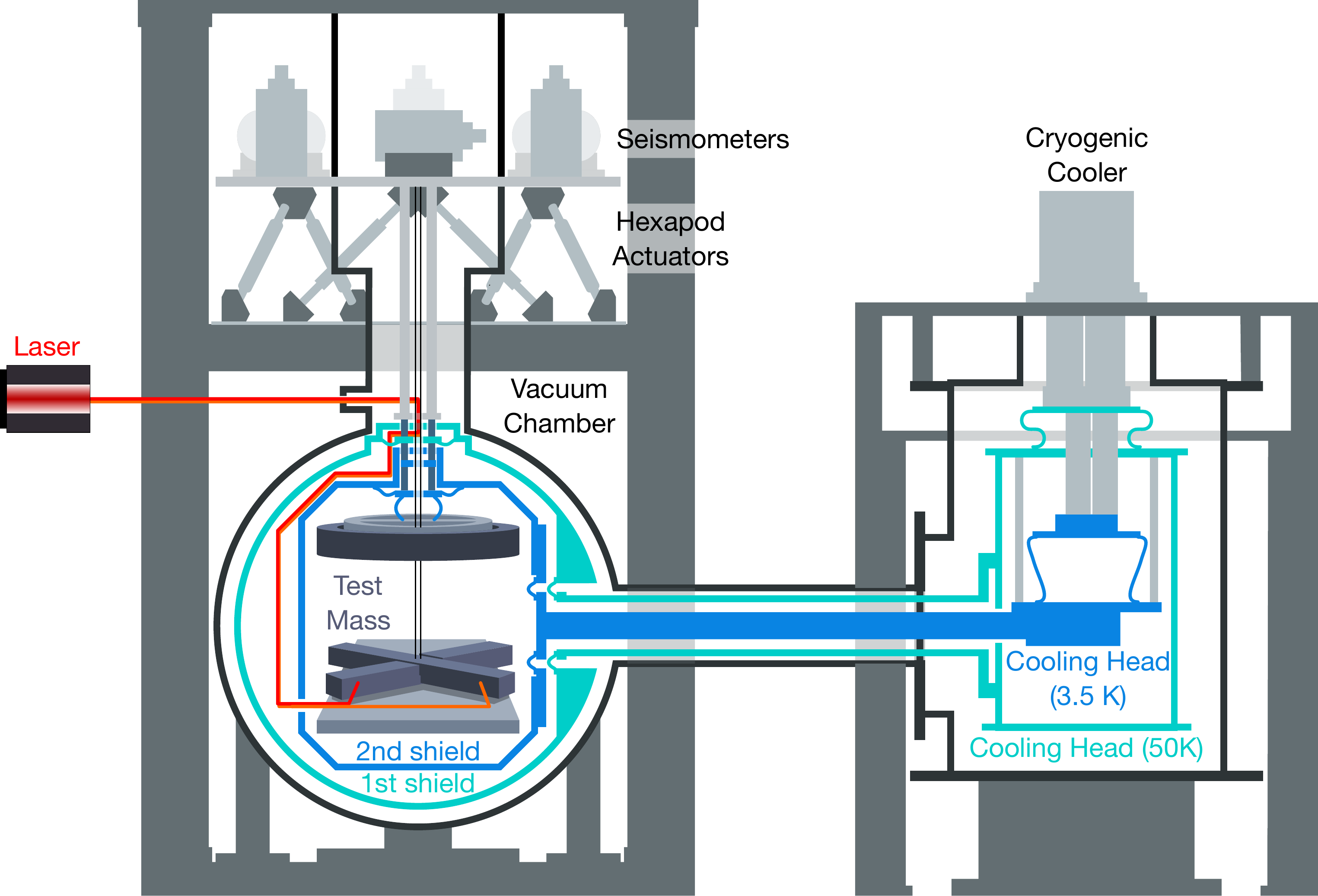}
    \caption{\label{Phase3_overview}
    Configuration of Phase-III TOBA.
    Inside a vacuum chamber, two radiation shields are connected to the cryocooler
    surrounding the suspension system.
    Two test mass bars are suspended by
    a wire with a high Q value.
    At the suspension point,
    an active vibration isolation system is implemented.
    The rotation of the bars are measured by
    an interferometer.
    }
\end{figure}

The configuration of Phase-III TOBA
is shown in Figure \ref{Phase3_overview}.
It consists of the following three parts:
\begin{itemize}
    \item A cryogenic suspension system;
    \item An active vibration isolation system;
    \item An interferometric readout system.
\end{itemize}

Here, we describe each system.

\subsubsection{\label{sec:4.1.1}
Cryogenic Suspension System}

The cryogenic system is one of the essential parts of Phase-III TOBA for reducing thermal noise.
The required temperature for Phase-III TOBA is
set to 4 K, and it is also desirable
to finish the cooling within an acceptable time
for operation.
Another important point is to suppress additional noises
introduced by the cooling system,
such as vibrations from \mbox{the cooler}.

The cooling system consists of
a cryocooler, two radiation shields,
and a main suspension system,
as shown in Figure \ref{Phase3_overview}.
A pulse tube cooler is used to cool
the radiation shields to $\SI{50}{K}$ (first shield)
and $\SI{3.5}{K}$ (second shield), respectively,
and the intermediate masses are connected
to the second shield via high-purity aluminum (6 N)
wires (heatlinks).
The test mass bars are cooled by thermal radiation
(>$\SI{100}{K}$)
and heat conduction through the suspension wires
(<$\SI{100}{K}$).
The suspension wires of the test mass bars are
made of silicon, which is known to have
low mechanical loss ($\phi<10^{-8}$)
at cryogenic temperatures~\cite{siliconflexure}.

Because each test mass is to be suspended by
a single silicon fiber,
the thermal exchange is not efficient and
the cooling time down to $\SI{4}{K}$
is approximately 14 days \cite{Shimoda_PhD}.
On the other hand, the suspension fibers
insulate the test mass bars well from the surroundings
and reduce other temperature fluctuations of the bars.

\subsubsection{\label{sec:4.1.2}
Active Vibration Isolation System}

To suppress the seismic vibration,
an active vibration isolation system
(AVIS) is implemented at the suspension point.

AVIS is a feedback system
that consists of multiple sensors and
hexapod actuators.
Six seismometers are arranged to monitor
all degrees of freedom of the suspension table.
The signals from the seismometers are
processed by a digital system, and
fed back to hexapod actuators,
which consist of
six piezoelectric actuators attached below the table.
The actuators are configured to actuate the table
in all degrees of freedom, including translation along three axes
and rotation around three axes.

AVIS is also utilized for suppressing the vibration of the cooler
introduced via the heatlinks attached to IMs.
Heatlinks on IMs are not directly connected to the second shield,
but first attached to a bypass stage,
which is connected to the suspension table via thermal insulation rods,
and then connected to the second shield.
By connecting the heatlinks on the stage,
the feedback system reduces
not only the seismic vibration,
but also vibrations transferred via the heatlinks.

The target performance of AVIS is
to reduce the vibration level to
below $10^{-7}\,\si{m/\sqrt{Hz}}$ at 0.1 Hz
for reducing the seismic cross-coupling noise
\cite{phase2-5},
and to below $10^{-10}\,\si{m/\sqrt{Hz}}$ at 1 Hz
to suppress nonlinear vibration noise \cite{nonlin_seis}.

\subsubsection{\label{sec:4.1.3}
Interferometric Readout System}

To achieve the target sensitivity,
we use Fabry--Pérot cavities to measure the motion
of the bars.
The surface of the test mass bar
is polished and coated
so that it works as mirrors,
to reduce the relative tilt of the mirrors
which introduces the cross-coupling.
The other mirrors, which form Fabry--Pérot cavities
with the surface of the bars,
are implemented on an optical bench.

One issue concerning the readout system
is the noise of the readout optics.
As the phase noise of the readout interferometer
limited the sensitivity of Phase-II TOBA,
the readout noise of the interferometer,
induced by ambient noise,
such as the vibration coupling and the temperature drift,
should be reduced sufficiently.
The required sensitivity for the readout noise
of the interferometer is set to
$\SI{6e-17}{m/\sqrt{Hz}}$ at 0.1 Hz.
To achieve this requirement,
a cryogenic-compatible monolithic interferometer
is planned to be introduced.

\subsection{\label{sec:4.2}
Target of Phase-III TOBA}

\begin{figure}[btp]
    \centering
    \includegraphics[width=1\linewidth]
    {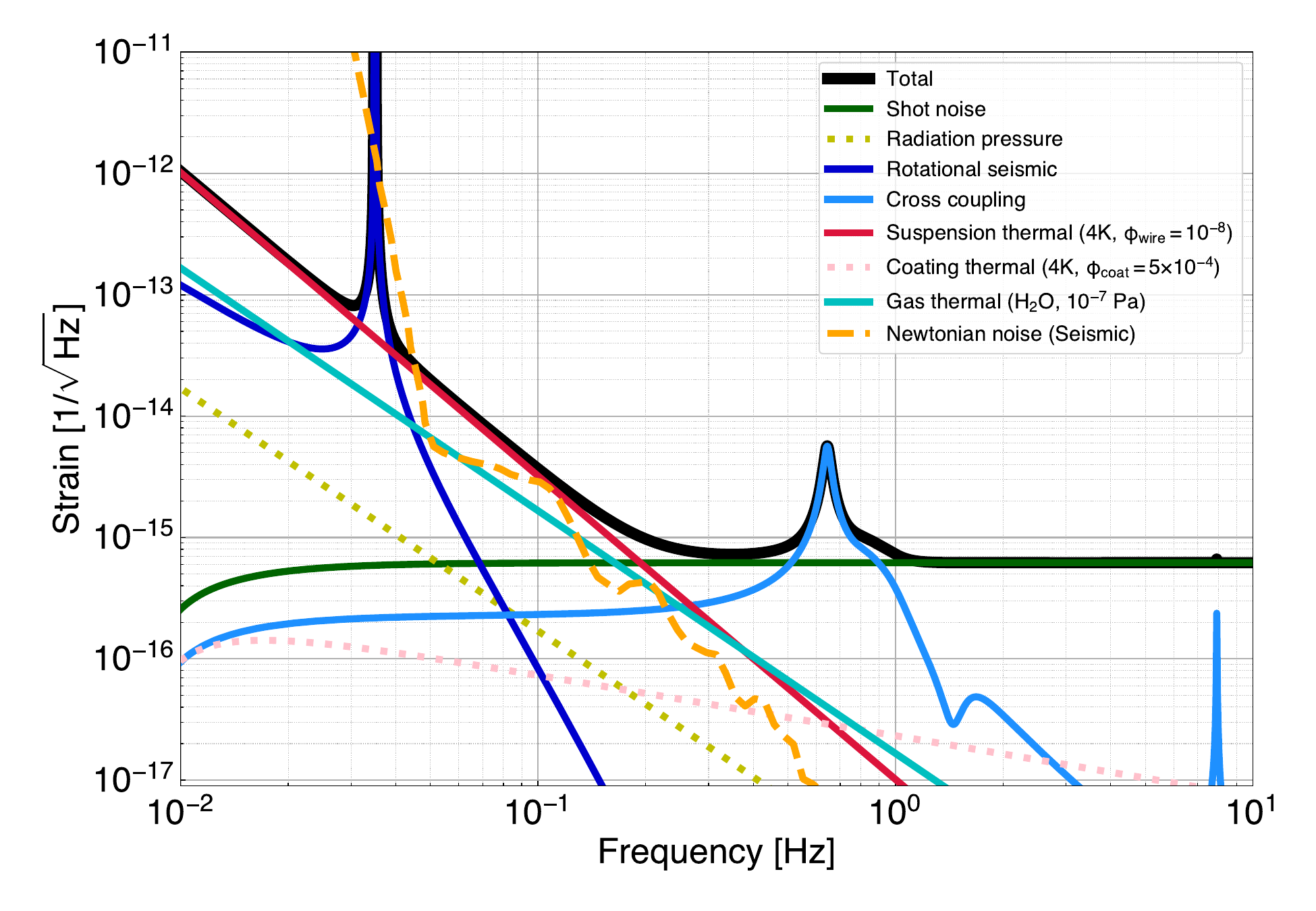}
    \caption{\label{Phase3_budget}
    The designed noise budget of Phase-III TOBA.
    The green line shows photon shot noise.
    The red line is the thermal noise of the suspension wire.
    The dark blue line and the sky blue line are
    the seismic noise from the ground rotation
    and the ground translation, respectively.
    The cyan line indicates the thermal noise
    from the residual water molecules.
    The black line is the sum of these noises.
    The orange dashed line indicates the estimated
    NN level of KAGRA site \cite{NN_KAGRA}
    to be measured with Phase-III TOBA.}
\end{figure}

With the systems mentioned above,
the design sensitivity of Phase-III TOBA
is shown in Figure \ref{Phase3_budget}.

As we mentioned in Section \ref{sec:2},
Phase-III TOBA has the potential
to detect earthquakes and measure NN directly
with a sensitivity of $10^{-15}/\si{\sqrt{Hz}}$.

\subsubsection{\label{sec:4.2.1}
Newtonian Noise Measurement}

The estimated level of NN is
in the order of $10^{-15}/\si{\sqrt{Hz}}$
around $\SI{0.1}{Hz}$.
Therefore, it is expected that
Phase-III TOBA can measure NN directly with the target sensitivity.
Although the frequency band
in which Phase-III TOBA can observe NN
is below the observation band of
the ground-based interferometric detectors,
measurement with Phase-III TOBA is useful
for testing the models of NN
and demonstrating its mitigation scheme.

\subsubsection{\label{sec:4.2.2}
Gravity Gradient Fluctuation
Induced by Earthquakes}

Previous research investigated the detectability
of earthquakes with Phase-III TOBA's target sensitivity
\cite{shimoda_warning,Shimoda_PhD}.
It has been shown that for earthquakes with a magnitude of $M_\mathrm{w} \,7$,
we can detect them $\sim$$\SI{10}{s}$ faster
than the current warning system
if they happen $\sim$$\SI{100}{km}$ away.
To identify the location of the center
of the earthquake with the same precision
as with the current system,
two detectors separated by $\sim$$\SI{75}{km}$
are necessary.

\subsection{\label{sec:4.3}
Current Achievements}

We have been developing Phase-III TOBA for years
and have achieved some of the goals
described in Section \ref{sec:4.1}.
We briefly introduce the current achievements
of these systems.

\subsubsection{\label{sec:4.3.1}
Cryogenic Suspension System}

\begin{figure}[btp]
    \centering
    \includegraphics[width=1\linewidth]
    {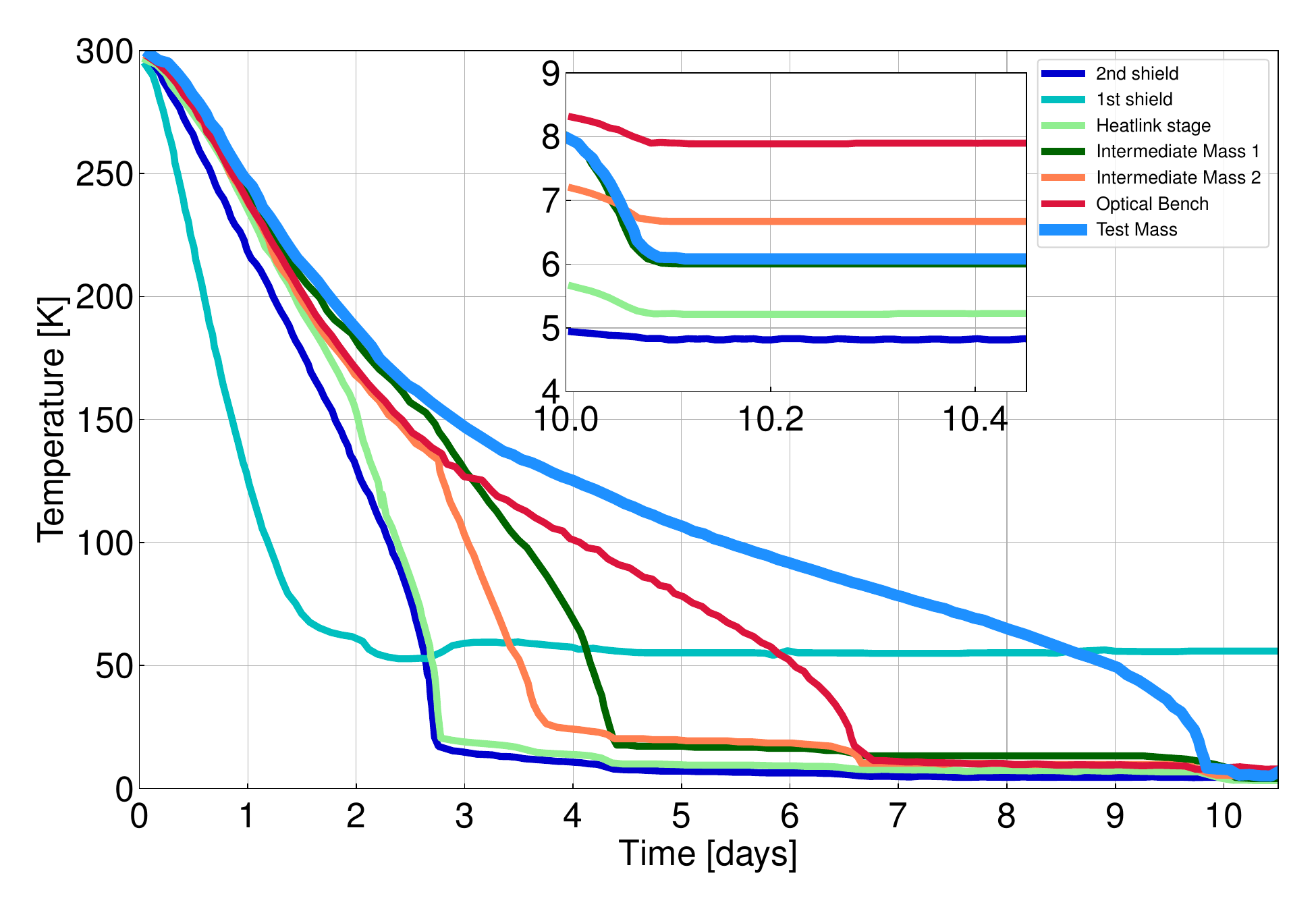}
    \caption{\label{cool_shimoda}
    The cooling curve of
    the cryogenic torsion pendulum
    measured in \cite{Shimoda_PhD}.
    The sky blue curve shows the temperature of
    the test mass bar.
    The test mass was cooled down to $\SI{6.1}{K}$
    in 10 days.
    }
\end{figure}

The basic cryogenic suspension system was
demonstrated in 2020 \cite{Shimoda_PhD}.
The measured cooling curve is shown in Figure \ref{cool_shimoda}.
The test mass bars were successfully cooled down
to 6.1 K in 10 days.
Because the silicon suspension wire is still under development,
in this demonstration, CuBe wires were used
for the suspension of the bars,
and 6N aluminum heatlinks were attached
between the intermediate mass and the test masses
to increase the heat conductance.

\subsubsection{\label{sec:4.3.2}
Active Vibration Isolation System}

\begin{figure}[btp]
    \centering
    \includegraphics[width=1\linewidth]{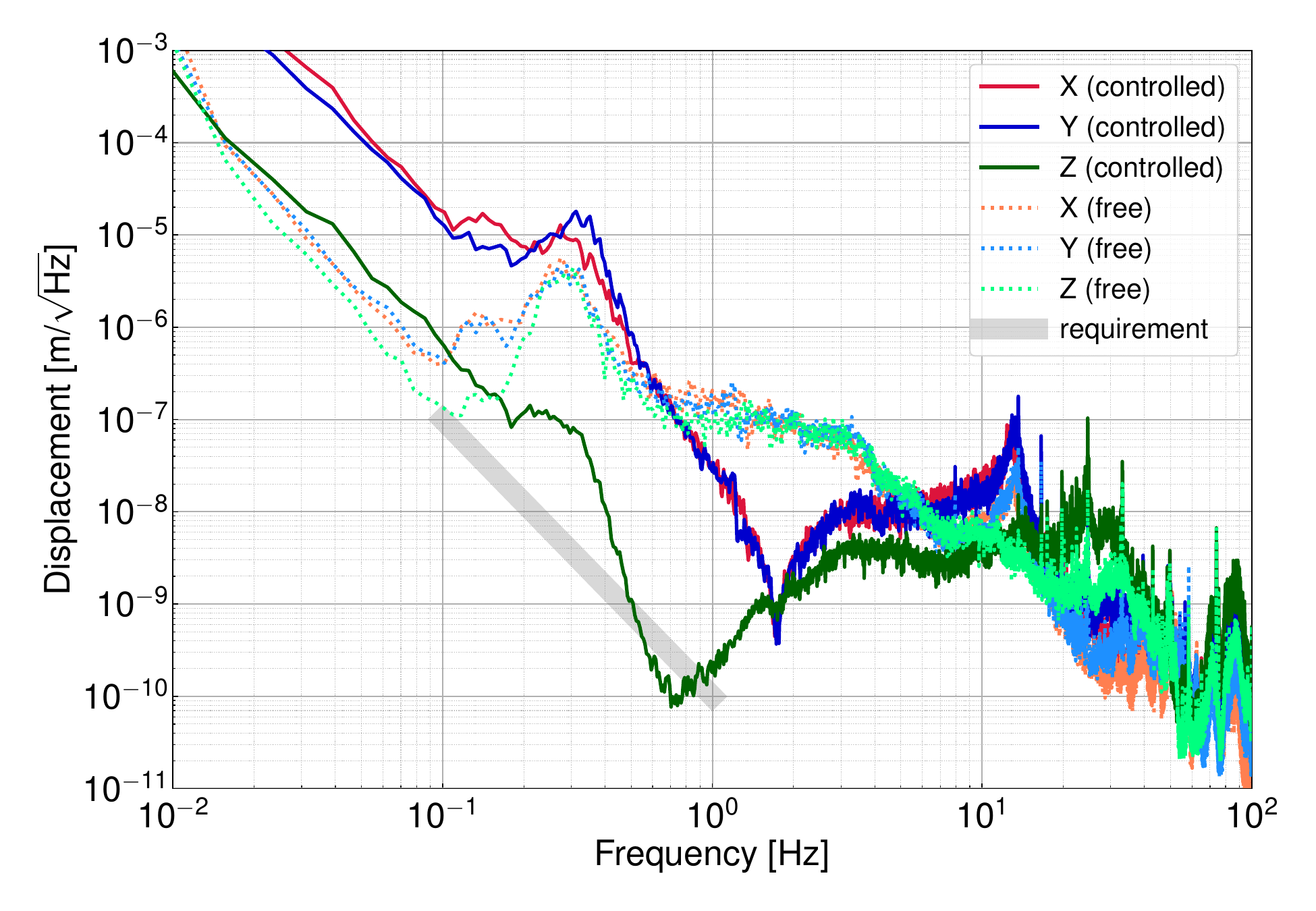}
    \caption{\label{AVIT_sens}
    The achieved performance of AVIS.
    The gray line indicates the requirement for AVIS.
    The red, blue, and green dashed lines are
    the vibration spectra without control
    in the $x$-, $y$-, and $z$-axes, respectively.
    The red, blue, and green solid lines show
    the vibration spectra with control of AVIS
    in the $x$-, $y$-, and $z$-axes, respectively.
    }
\end{figure}

A prototype of the vibration isolation system
was developed and the performance was tested in 2019 without the suspension system.
The achieved performance is shown in Figure \ref{AVIT_sens}.
The vertical seismic vibration was suppressed
by $10^3$ around 0.7 Hz
and the horizontal vibration by $\num{3e-2}$ around 1.7 Hz.
The performance was still not sufficient to meet the requirement.
One reason was tilt-horizontal coupling
\cite{tilt_coupling}
in frequencies below $\SI{0.5}{Hz}$,
which smeared the true signal of the horizontal motion
and introduced instability in the control.
Another reason was the parasitic resonance modes of the
supporting frame of the system,
which prevented us from increasing the feedback gain
and resulted in an insufficient suppression ratio.

\subsubsection{\label{sec:4.3.3}
Interferometric Readout System}

\begin{figure}[btp]
    \centering
    \includegraphics[width=1\linewidth]{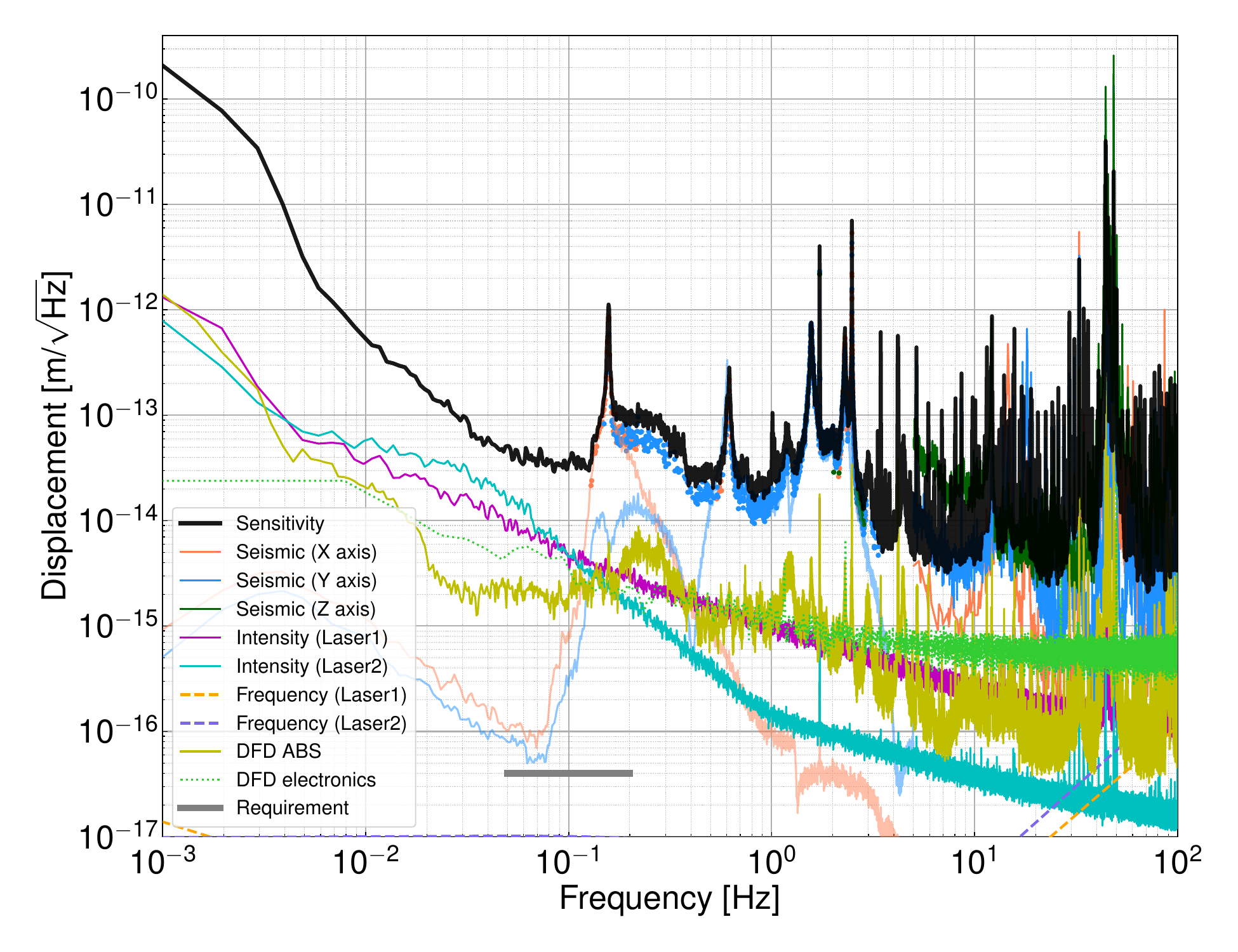}
    \caption{\label{mono_sens}
    The achieved sensitivity of
    the cryogenic monolithic interferometer \cite{Takano_PhD}.
    The sensitivity shown as the black curve is
    limited by seismic noise (red in the $x$-axis,
    blue in the $y$-axis, and green in the $z$-axis).
    The gray line indicates the requirement for
    the monolithic interferometer.
    }
\end{figure}

\begin{sloppypar}
    The cryogenic monolithic interferometer
    was demonstrated in 2024 \cite{Takano_PhD}.
    In this demonstration, the readout optics made of silicon
    were glued on a silicon breadboard
    and consisted of a monolithic interferometer
    with one test mass bar fixed on the breadboard as well.
    The interferometer was operated stably at 12 K
    and achieved the sensitivity of
    $\SI{3.6e-14}{m/\sqrt{Hz}}$ at 0.1 Hz.
    The performance was limited by the seismic noise
    on the horizontal axis.
    The sensitivity and the estimated noise contributions
    are shown in \mbox{Figure \ref{mono_sens}}.
\end{sloppypar}

Another method to measure the rotation of the bar,
the Cavity-Amplified Angular Sensor (CAAS),
was also proposed \cite{CWFS}.
CAAS utilizes an optical cavity for the amplification of the angular signal,
and has better sensitivity than
normal angular measurement methods, such as optical levers.
Compared to the rotation measurements by
differential Fabry--Pérot cavities,
direct rotation measurement by CAAS has the advantage
of smaller cross-coupling noise.
So far, the basic principle has been demonstrated \cite{CWFS}
and further improvements \mbox{are ongoing}.

\subsection{\label{sec:4.3.4}Beyond Phase-III TOBA}

As we see in the previous section,
some of the techniques necessary for Phase-III TOBA
have been demonstrated.
Although they still need further improvement,
the outlook toward the Final TOBA
should be discussed in parallel.

\begin{sloppypar}
    To scale the detector up to $\SI{10}{m}$ in length,
    there will be many difficulties.
    For example, suspending such gigantic objects
    with a single fiber requires special attention to the mechanical design,
    such as tuning resonant frequency
    while maintaining low mechanical loss
    and sufficient tensile strength.
    Cooling such a enormous system is
    another potential issue.
    Solving such technical issues will require
    additional study and substantial costs,
    which could diminish its advantage
    over space detectors.
\end{sloppypar}

Nevertheless, TOBA has its own valuable scientific targets,
such as NN measurement and earthquake detection,
which cannot be observed with space detectors.
These targets are expected to be measured
with Phase-III TOBA
if all the technical achievements are realized,
and future TOBA setups will obtain
more precise results.

\section{\label{sec:5}Conclusions}

In this paper, an overview of TOBA is presented.
The final target of TOBA is described,
and the current status of the technical demonstration is provided.

A small prototype named Phase-III TOBA is under development
to achieve the sensitivity of $10^{-15}\,/\si{\sqrt{Hz}}$.
With this sensitivity, geophysical targets
such as earthquakes and NN
are expected to be detectable.
To achieve the target sensitivity,
technical demonstrations of each subsystem are ongoing.

After achieving the target of Phase-III TOBA,
we will move on to building a large-scale detector
to achieve the sensitivity of $10^{-19}\,/\si{\sqrt{Hz}}$.
This detector would enable us to observe GWs
from the merger of IMBH binaries.

\vspace{16pt}
\noindent
\textbf{Author Contributions}:
Conceptualization, S.T., T.S., and M.A.;
methodology, S.T., T.S., Y.O., C.P.O., and M.C.;
software, S.T. and T.S.;
validation, K.K., Y.M., and M.A.;
formal analysis, S.T. and T.S.;
investigation, S.T., T.S., and M.A.;
resources, S.T. and T.S.;
data curation, M.A.;
writing—original draft preparation, S.T.;
writing—review and editing, S.T., C.P.O., P.W.F.F.,
Y.O., K.K., N.K., T.S., and M.A.;
visualization, S.T.;
supervision, M.A.;
project administration, M.A.;
funding acquisition, all authors.
All authors have read and agreed to the published version of the manuscript.

\noindent
\textbf{Fundiwding}:
JSPS KAKENHI, Grants Number JP16H03972, JP24244031,
JP18684005, and JP22H01246.
This work was also supported by
the MEXT Quantum Leap Flagship Program (MEXT Q-LEAP),
Grant Number JPMXS0118070351.

\noindent
\textbf{Data Availability Statement}:
Dataset available on request from the authors.

\noindent
\textbf{Acknowledgments}:
We would like to thank Takafumi Ushiba (ICRR)
for helpful discussion about cryogenics.
We also would like to thank Yoichi Aso (NAOJ)
for offering materials for the cryogenic experiments.
Finally, we would like to acknowledge
the great support of Shigemi Otsuka
and Togo Shimozawa (The University of Tokyo)
for making the mechanical parts in the experiments.

\noindent
\textbf{Conflicts of Interest}:
The authors declare no conflicts of interest.

\section{\label{sec:A}
Derivation of the Response of a Torsion Pendulum
to Gravitational Waves}

In this section, we derive the mechanical response
of a torsion pendulum to gravitational waves.

\subsection{\label{sec:A.1}
Torque Induced by Gravitational Waves}

Consider a GW coming along the $z$-axis
described by Equations \eqref{GW_z} and \eqref{GW_pol}.
Within a region whose scale is small enough
compared to the wavelength of the GW,
the force $dF_\mathrm{GW}^i$ induced by the GW
acting on a particle with mass $m$ at $x^i$
is given by \cite{GW_maggiore}
\begin{equation}
    dF_\mathrm{GW}^i = \frac{1}{2}
    m\ddot{h}^i_jx^j.
    \label{dF_GW}
\end{equation}

Here, $h_{\mu\nu}$ ($\mu,\nu\in\{0,1,2,3\}$
are the small deviations of the metric tensor
$g_{\mu\nu}$ from flat space-time
$\eta_{\mu\nu}=\mathrm{diag}(-1,+1,+1,+1)$,
and $h_{ij}$ ($i,j\in{1,2,3}$)
are the spatial components of $h_{\mu\nu}$.
Up to linear order of $h_{\mu\nu}$,
the indices of $h_{\mu\nu}$ are raised by $\eta^{\mu\nu}$.
In particular, $h^i_j$ are \mbox{given by}
\begin{equation}
    h^i_j = \eta^{i\alpha}h_{\alpha j}
    = \delta^{ik} h_{k j}
    = h_{ij}.
\end{equation}

Next, consider a bar located along the $x$-axis,
as shown in Figure \ref{TOBA}.
Using \mbox{Equation \eqref{dF_GW}},
the force induced by the GW
acting on a discrete element of the bar $dV$ at $\xi^i$
is given by
\begin{equation}
	dF_\mathrm{GW}^i = \frac{1}{2}
    \rho dV\ddot{h}^i_j\xi^j,
    \label{df_bar}
\end{equation}
where $\rho$ is the density of the bar and
$\xi^i$ is the location of $dV$,
whose origin is on the center of mass of the bar.
The energy stored in $dV$ by $dF_\mathrm{GW}$
is given by integrating $dF_\mathrm{GW}$
along $\xi^i$,
\begin{equation}
    dU_\mathrm{GW} =
    -\int_0^{\xi^i}d\xi'_idF_\mathrm{GW}^i
    = \int_0^{\xi^i}d\xi'_i
    \frac{1}{2}\rho dV\ddot{h}^i_j\xi^j
    = \frac{1}{4}\rho dV\ddot{h}_{ij}\xi^i\xi^j.
\end{equation}

Then, integrating $dU_\mathrm{GW}$ over
all the volume of the bar,
we obtain the total energy stored in the bar
$U_\mathrm{GW}$ by
\begin{equation}
    U_\mathrm{GW} = \int_{V}dU_\mathrm{GW}
    = \frac{1}{4}\ddot{h}_{ij}
    \int_{V}dV\rho \xi_i\xi_j.
\end{equation}

A torque on the bar is the rate of change
of the energy
related to the rotation $\theta$.
That is,
\begin{equation}
    N_\mathrm{GW} =
    -\frac{\partial U_\mathrm{GW}}
    {\partial\theta}
    = \frac{1}{4}\ddot{h}_{ij}q^{ij}.
    \label{N_GW0}
\end{equation}

Here, $q^{ij}$ is the quadrupole moment of the bar
defined as
\begin{equation}
    q^{ij} = \int_{V}dV\rho
    \left(
    \xi^i w^j + \xi^j w^i -
    \frac{2}{3}\delta^{ij}\xi^k w_k
    \right),
    \label{quad_mom}
\end{equation}
where $w^i$ is the mode function
and in the case of the rotation around $z$-axis,
\begin{equation}
    w^i =
    \begin{pmatrix}
        -y \\ x \\ 0
    \end{pmatrix}.
    \label{mode_func}
\end{equation}

Substituting Equation \eqref{mode_func}
into Equation \eqref{quad_mom},
we obtain the components of the quadrupole moment as follows:
\begin{align}
    q_+ &:= q_{11} = -q_{22} =
    -\int_{V}dV\rho 2xy,
    \label{q_+} \\
    q_{\times} &:= q_{12} = q_{21} =
    \int_{V}dV\rho(x^2-y^2).
    \label{q_x}
\end{align}

Using Equations \eqref{N_GW0}, \eqref{q_+},
and \eqref{q_x},
the torque on the bar induced by GWs is \mbox{written as}
\begin{equation}
    N_\mathrm{GW} =
    \frac{\ddot{h}_{11}-\ddot{h}_{22}}{4}q_+
    + \frac{1}{2}\ddot{h}_{12}q_\times.
    \label{N_GW1}
\end{equation}

Substituting Equations \eqref{GW_z} and \eqref{GW_pol} into Equation \eqref{N_GW1}, we obtain
\begin{equation}
    N_\mathrm{GW} =
    \frac{\ddot{h}_+}{2}q_+ +
    \frac{\ddot{h}_\times}{2}q_\times.
    \label{N_GW2}
\end{equation}

\subsection{\label{sec:A.2}
Mechanical Response of a Torsion Pendulum}

We consider the frequency response of the torsion pendulum.
The equation of motion about $\theta$ is given by
\begin{equation}
    I\ddot{\theta}(t) + \kappa\theta(t) =
    N_\mathrm{GW}(t).
    \label{EOM_t}
\end{equation}

Here, $I$ is the moment of inertia of the bar
around the rotational axis ($z$-axis), and
$\kappa$ is the spring constant
of the suspension wire
in the torsional mode.
Applying Fourier transformation to Equation \eqref{EOM_t},
we obtain
\begin{equation}
    \frac{\theta(f)}{N_\mathrm{GW}(f)} =
    \frac{1}
    {\kappa(1+i\phi_\mathrm{rot})+(2\pi f)^2I}
    = \frac{1}
    {4\pi^2I(f_0^2(1+\phi_\mathrm{rot})+f^2},
    \label{theta_f}
\end{equation}
where
\begin{equation}
    f_0 := \frac{1}{2\pi}
    \sqrt{\frac{\kappa}{I}}
\end{equation}
is the resonant frequency of the rotation mode.
Here, the loss angle $\phi_\mathrm{rot}$
is the imaginary part
of the spring constant $\kappa$,
which is introduced to describe
mechanical energy loss of the suspension wire.
Plugging Equation \eqref{N_GW2} into Equation \eqref{theta_f},
we obtain
\begin{align}
    \theta(f) &=
    \left(
        \frac{q_+}{2I}h_+(f) +
        \frac{q_\times}{2I}h_\times(f)
    \right)
    \frac{f^2}{f_0^2(1+i\phi_\mathrm{rot})-f^2}
    \\
    &= \sum_{A} H_A(f)h_A(f)
\end{align}
where $A=\{+, \times\}$ and
\begin{equation}
    H_A(f) := \frac{q_A}{2I}
    \frac{f^2}{f_0^2(1+i\phi_\mathrm{rot})-f^2}.
\end{equation}

Next, consider the relation
between the moment of inertia $I$
and the quadrupole moments $q_+$ and $q_\times$.
If the bar shape is a rectangular cuboid
with mass $M$ and
length along each axis $l_x$ and $l_y$,
we can write $I, q_+$, and $q_\times$ as
\begin{align}
    I &= \int_VdV\rho(x^2+y^2)
    = \frac{1}{12}M(l_x^2+l_y^2), \\
    q_+ &= \int_VdV\rho(-2xy)
    = 0, \\
    q_\times &= \int_VdV\rho(x^2-y^2)
    = \frac{1}{12}M(l_x^2-l_y^2)
    \label{qx}
\end{align}

If the bar is along the $x$-axis
and the aspect ratio is large enough,
$l_x\gg l_y$, we can approximate
$q_\times/I\simeq1$, and we obtain
\begin{equation}
    H_+(f) = 0,\,\,
    H_\times(f) \simeq \frac{1}{2}
    \frac{f^2}{f_0^2(1+i\phi_\mathrm{rot})-f^2}.
\end{equation}

\bibliography{ref}

\end{document}